\documentclass[12pt]{article}
\usepackage{fullpage}
\usepackage{epsf}
\usepackage{epsfig}
\usepackage{amsthm}
\usepackage{amsfonts}
\usepackage{color}
\usepackage{amsmath}
\usepackage{setspace}
\usepackage{natbib}
\usepackage{url}

\begin{document}

\newcommand{\bm}[1]{\mbox{\boldmath $#1$}}
\newcommand{\mb}[1]{\mathbf{#1}}
\newcommand{\bE}[0]{\mathbb{E}}
\newcommand{\bP}[0]{\mathbb{P}}
\newcommand{\ve}[0]{\varepsilon}
\newcommand{\Var}[0]{\mathbb{V}\mathrm{ar}}
\newcommand{\Corr}[0]{\mathbb{C}\mathrm{orr}}
\newcommand{\Cov}[0]{\mathbb{C}\mathrm{ov}}
\newcommand{\mN}[0]{\mathcal{N}}
\newcommand{\iidsim}[0]{\stackrel{\mathrm{iid}}{\sim}}
\newcommand{\NA}[0]{{\tt NA}}
\newcommand{\argmax}{\operatornamewithlimits{argmax}}

\newcommand{\jarad}[1]{{\color{red}JARAD: #1}}
\newcommand{\bobby}[1]{{\color{blue}BOBBY: #1}}

\title{\vspace{-1cm}Massively parallel\\ approximate Gaussian process regression}
\author{Robert B.~Gramacy\\
  Booth School of Business\\
  The University of Chicago\\
    {\tt rbgramacy@chicagobooth.edu}
  \and
  Jarad Niemi\\
  Department of Statistics\\
  Iowa State University\\
  {\tt niemi@iastate.edu}
  \and
  Robin M.~Weiss\\
  Research Computing Center\\
  The University of Chicago\\
  {\tt robinweiss@uchicago.edu}
}
 \date{}

\maketitle

%\doublespacing
\vspace{-0.2cm}
\begin{abstract}
  We explore how the big-three computing paradigms---symmetric multi-processor
  (SMP), graphical processing units (GPUs), and cluster computing---can
  together be brought to bear on large-data Gaussian processes (GP) regression
  problems via a careful implementation of a newly developed local
  approximation scheme.  Our methodological contribution focuses primarily on
  GPU computation, as this requires the most care and also provides the
  largest performance boost. However, in our empirical work we study
  the relative merits of all three paradigms to determine how best to combine
  them. The paper concludes with two case studies.  One is a real data
  fluid-dynamics computer experiment which benefits from the local nature of
  our approximation; the second is a synthetic example designed to find
  the largest data set for which (accurate) GP emulation can be performed on a
  commensurate predictive set in under an hour.

  \bigskip
  \noindent {\bf Key words:} emulator, nonparametric regression, graphical
  processing unit, symmetric multi-processor, cluster computing, big data,
  computer experiment
\end{abstract}

\section{Introduction}
\label{sec:intro}

Gaussian processes \cite[see, e.g.,][]{rasmu:will:2006} form the canonical
regression model for data arising from computer experiments
\citep{sant:will:notz:2003}.  Their nonparametric flexibility, interpolative
capability, and conditionally analytic predictive distributions with high accuracy
and appropriate coverage make them ideally suited to applications where
accuracy, and full uncertainty quantification/propagation, are equally
important. Some examples include design
\citep{sant:will:notz:2003}, sequential design \citep{seo:etal:2000}, optimization 
\citep{jones:schonlau:welch:1998}, contour finding \citep{ranjan:haynes:karsten:2011},  
and calibration \citep{kennedy:ohagan:2001,bayarri:etal:2007}, to name just a few.

The main disadvantage to Gaussian process (GP) regression models is
computational.  Inference and prediction require decomposing an $N \times N$
matrix, for $N$ observations, and that usually requires an $O(N^3)$ operation.
With modern desktop computers, that limits GPs to $N$ in the low thousands for
point inference (e.g., via maximum likelihood estimation [MLE] or cross
validation [CV]), and to the hundreds for sampling methods (Bayesian Monte
Carlo or bootstrap). Point inference can be pushed to $N$ in the tens of
thousands when modern supercomputer computing facilities are paired with new
distributed linear algebra libraries, as illustrated by
\cite{paciorek:etal:2013}.

As data sets become ever larger, research into approximate GP modeling has
become ever more frenzied.  Early examples include the works of
\cite{vecchia:1988,nychka:wykle:royle:2002,stein:chi:welty:2004,qc:rasmu:2005,furrer:genton:nychka:2006,cressie:joh:2008}.  More recent examples include those of
\cite{haaland:qian:2012,sang:huang:2012,kaufman:etal:2012,eidsvik2013estimation}.  
Sparsity is a recurring theme.  The approximations involve either
explicitly working with a subset of the data, or imposing a covariance structure
which produces sparse matrices that can be quickly decomposed.  Another way
to get fast inference is to impose structure on the design, e.g., forcing a
lattice design. This can lead to substantial shortcuts in the calculations
\citep[e.g.,][]{gilboa:etal:2012,plumlee:2013}, but somewhat limits applicability.

In this work we consider the particular approximation suggested by
\cite{gramacy:apley:2014}.  That approximation drew inspiration from several
of the works referenced above, but primarily involves modernizing an old idea
of local kriging neighborhoods \citep[][pp.~131--134]{cressie:1993} by
borrowing from active learning heuristics for sequential design
\citep{seo:etal:2000}.  The idea is to build a GP
predictor for a particular location, $x$, by greedily building a sub-design $X_n(x)
\subseteq X$, where $n \ll N$, according to an appropriate criteria. Then,
prediction over a vast grid can be parallelized by processing each element, $x$,
of the grid independently of the others.  Such independence can also yield a
thrifty nonstationary modeling feature.

Our primary contribution centers around recognizing that the criteria, which
must be repeatedly evaluated over (nearly) the entire design space $X$, can be
implemented on a graphical processing unit (GPU).  In essence, we are
proposing to nest a parallel (GPU) computation within an already parallelized
symmetric multiprocessor (SMP) environment.  Both the GPU implementation (in
{\tt CUDA}), and its interface to the outer parallel scheme (via {\tt
OpenMP}), must be treated delicately to be efficient. We then round out the
supercomputing trifecta by distributing computation on a cluster of multi-core,
and multi-GPU, nodes.

GPUs offer great promise in scientific computing, in some cases leading to
100x speedups.  We find more modest speedups in our examples (40-60x), echoing similar results obtained with GPU-accellerated large matrix operations
\citep{franey:ranjan:chipman:2012,eidsvik2013estimation,paciorek:etal:2013}.  
In contrast to these works, however, we do not make use of library routines.
In fact, our approximations explicitly keep the required matrices small. Our
repetitive local searches involve low-level operations which can be
implemented on the GPU with a very small (and completely open source) kernel.
The entire implementation, including {\tt CUDA}, {\sf C}, and {\sf R}
subroutines, is made available as an {\sf R} package called {\tt laGP}
\citep{laGP}.

The remainder of the paper is outlined as follows.  Section \ref{sec:review}
reviews GP computer modeling generally, and the \cite{gramacy:apley:2014}
local approximate GP scheme specifically, with focus on the particular
subroutine that is re-implemented in this paper.  Section
\ref{sec:gpu}  details  our {\tt CUDA} implementation of that subroutine, and
in Section \ref{sec:results} we study the speedups obtained in isolation
(i.e., compared to a CPU version of the same subroutine), and within the wider
context (incorporating the calling environment on a multi-core, multi-GPU
compute node) on a toy problem.  In Section \ref{sec:big} we augment with an
off-the-shelf, simple network of workstations (SNOW), cluster computing
facility in order to emulate a large real-data computer experiment from the
literature.  We then separately consider a synthetic data-generating mechanism
to find the largest problem we can solve with an hour of (multi-node cluster)
computing time. The paper concludes with a brief discussion in Section
\ref{sec:discuss}.

\section{Computer model emulation}
\label{sec:review}

Computer simulation of a system under varying conditions represents a
comparatively inexpensive alternative to actual physical experimentation
and/or monitoring.  Examples include aeronautics
(designing a new aircraft wing) and climate science (collecting
atmospheric ozone data).  In some cases it is the only (ethical) alternative, e.g.,
in epidemiology. Still, computer models can be complex and computationally
demanding, and therefore only a limited (if still vast) number of runs $D_N =
(x_1, y_1), \dots, (x_N, y_N)$ for input conditions $x_i$, producing outputs
$y_i$, can be obtained.  Computer model {\em emulation}, therefore, remains an
integral component of many applications involving data arising from computer
simulation.  Given the data $D_N$, an emulator provides a distribution over
possible responses $Y(x)|D_N$ for new inputs $x$. A key requirement is that
the emulator be able to provide that distribution at much lower
computational expense than running new simulations.

\subsection{Gaussian process regression}
\label{sec:gp}

The GP regression model is canonical for emulation, 
% for many applications 
primarily for the following two reasons.
\begin{enumerate}
\item The predictive equations $p(y(x)|D_N, K_\theta)$ have a closed form
given a small number of ``tuning'' parameters, $\theta$, describing the
correlation structure $K_\theta(\cdot,\cdot)$, which is discussed separately
below.  They are Student-$t$ with degrees of freedom $N$,
\begin{align} 
  \mbox{mean} && \mu(x|D_N, \theta) &= k^\top(x)  K^{-1}Y,
\label{eq:predgp} \\ 
\mbox{and scale} && 
 \sigma^2(x|D_N, \theta) &=   
\frac{\psi [K(x, x) - k^\top(x)K^{-1} k(x)]}{N},
\label{eq:preds2}
\end{align}
where $k^\top(x)$ is the $N$-vector whose $i^{\mbox{\tiny th}}$ component is
$K_\theta(x,x_i)$, $K$ is an $N\times N$ matrix whose entries are
$K_\theta(x_i, x_j)$, and $\psi = Y^\top K^{-1} Y$.  Using properties of the
Student-$t$, the variance of $Y(x)$ is $V(x) \equiv \Var[Y(x)|D_N,\theta] =
\sigma^2(x|D_N,\theta)\times N/(N - 2)$.

Observe that the mean is a linear predictor, which depends on the responses
$Y$, and that the variance is independent of $Y$ given $K_\theta(\cdot,
\cdot)$.  The result is a ``football-shaped'' predictive interval which is
wide away from data locations $x_i$, and narrows 
at the $x_i$---a visually appealing feature for an emulator.

\item Maximum likelihood inference for $\theta$ is straightforward given
analytic forms of the (marginalized) likelihood equations 
\begin{equation}
p(Y|\theta) = \frac{\Gamma[N/2]}{(2\pi)^{N/2}|K|^{1/2}} \times
\left(\frac{\psi}{2}\right)^{\!-\frac{N}{2}},
\label{eq:gpk}
\end{equation}
whose derivatives, for Newton-like optimization, are also available
analytically.
\end{enumerate}
Together, analytic prediction and
straightforward optimization for inference, make for a relatively easy
implementation of a non-parametric regression.  %% Open source codes abound.

The choice of correlation structure, $K_\theta(\cdot, \cdot)$, can have a
substantial impact on the nature of inference and prediction, restricting the
smoothness of the functions and controlling a myriad of other aspects. However
there are several simple default choices that are popular in the literature.
In this paper we use an isotropic Gaussian correlation $K_{\theta,\eta}(x,x')
=
\exp\{ - ||x - x'||^2/\theta\}$, where $\theta$ is called the {\em
lengthscale} parameter.  We make this choice purely for simplicity of the
exposition, and because it is historically the most common choice for computer
experiments.  The general methodology we present is independent of this
choice.

Unfortunately, the above equations reveal a computational expense that depends
on the size of the correlation matrix, $K$. Inverse and determinant
calculations are $O(N^3)$ which, even for modest $N$, can mean that (in spite
of the above attractive features) GPs may not satisfy the key requirement of
being fast relative to the computer simulation being emulated. Advances
in hardware design, e.g., multi-core machines and GPUs, may offer some
salvation.  Recently several authors
\citep{franey:ranjan:chipman:2012,eidsvik2013estimation,paciorek:etal:2013}
have described custom GP prediction and inference schemes which show a
potential to handle much larger problems than ever before.

\subsection{Local approximate Gaussian process modeling}
\label{sec:laGP}

It makes sense to develop emulators which can exploit these new resources,
especially as they move into the mainstream.  For obvious reasons, emulation
in better than $O(N^3)$ time is also desirable, and for that imposing sparsity is
a popular tactic.
 \cite{gramacy:apley:2014} proposed a local scheme leveraging sparsity  towards 
 providing fast and accurate prediction, ideal for computer model
 emulation on modern multi-core desktops.

The idea is to focus, specifically, on the prediction problem at a particular
location, $x$.  \citeauthor{gramacy:apley:2014} recognized, as many others
have before, that data with inputs far from $x$ have vanishingly small
influence on the resulting GP predictor (assuming typical distance-based
correlation functions). Exploiting that, the scheme seeks to build a GP
predictor from data $D_n(x) \equiv D_n(X_n(x))$ obtained on a sub-design
$X_n(x)$ of the full design $X \equiv X_N$, where $n
\ll N$.  One option is a so-called {\em nearest neighbor} (NN) sub-design, where
$D_n$ is comprised of the inputs in $X$ which are closest to $x$, measured
relative to the chosen correlation function, but this is known to be sub-optimal
\citep{vecchia:1988}. It is better to take at least a few design
points farther away in order to obtain good estimates of the parameter
$\theta$ \citep{stein:chi:welty:2004}.  However, searching for the optimal
design $\hat{D}_n(x)$, according to almost any criteria, is a combinatorially
huge undertaking.  The interesting pragmatic research question that remains
is: is it possible to do better than the NN scheme without much extra
computational effort?

\citeauthor{gramacy:apley:2014} demonstrated that it is indeed possible, with the
following greedy scheme.  Suppose a local design $X_j(x)$, $j<n$, has been built-up already,
and that a GP predictor has been inferred from data $D_j(x)$.
Then, choose $x_{j+1}$ by searching
amongst the remaining unchosen design candidates $X_N \!\setminus\! X_j(x)$
according to a criterion, discussed momentarily.  Augment the
data set $D_{j+1}(x) = D_j \cup (x_{j+1}, y(x_{j+1}))$ to include the chosen
design point and its corresponding response, and update the GP predictor.
Updating a GP predictor is possible in $O(j^2)$ time
\citep{gramacy:polson:2011} with judicious application of the partitioned
inverse equations \citep{barnett:1979}.  So as long as each search for
$x_{j+1}$ is fast, and involves no new operations larger than $O(j^2)$, then
the final scheme, repeating for $j=n_0, \dots, n$ will require $O(n^3)$
time, just like the NN scheme.

\citeauthor{gramacy:apley:2014} considered two criteria in addition to NN, one being
a special case of the other.  The first is to minimize the empirical Bayes
mean-square prediction error (MSPE): $ J(x_{j+1}, x) = \bE\{ [Y(x) -
\mu_{j+1}(x|D_{j+1}, \hat{\theta}_{j+1})]^2 | D_j(x) \}$ where $\hat{\theta}_{j+1}$ is the estimate for $\theta$ based on $D_{j+1}$.  The predictive mean
$\mu_{j+1}(x|D_{j+1}, \hat{\theta}_{j+1})$ follows equation \eqref{eq:predgp}, except
that the $j\!+\!1$ subscript has been added in order to indicate dependence on
$x_{j+1}$ and the future, unknown $y_{j+1}$. They then derive the approximation
\begin{equation}
J(x_{j+1},x) \approx  \left. V_j(x | x_{j+1}; \hat{\theta}_j) + \left(\frac{\partial \mu_j(x;
    \theta)}{\partial \theta}
\Big{|}_{\theta = \hat{\theta}_j}\right)^2 \right/
\mathcal{G}_{j+1}(\hat{\theta}_j).
\label{eq:mspe}
\end{equation}
The first term in (\ref{eq:mspe}) estimates predictive variance at $x$ after $x_{j+1}$ is
added into the design,
\begin{align}
V_j(x|x_{j+1}; \theta) &= \frac{(j+1)\psi_j}{j(j-1)} v_{j+1}(x; \theta),
\nonumber \\ 
\mbox{where } \;\;\; v_{j+1}(x; \theta) &= \left[ K_{j+1}(x, x) -
k_{j+1}^\top(x) K_{j+1}^{-1} k_{j+1}(x) \right].
\label{eq:newv}
\end{align}
Minimizing predictive variance at $x$ is a sensible goal.  The second term in
(\ref{eq:mspe}) estimates the rate of change of the predictive mean at $x$,
weighted by the expected {\em future} inverse information,
$\mathcal{G}_{j+1}(\hat{\theta}_j)$, after $x_{j+1}$ and the corresponding
$y_{j+1}$ are added into the design.  Note that this weight does not depend on
$x$, but in weighting the rate of change (derivative) of the predictive mean
at $x$ it is ``commenting'' on the value of $x_{j+1}$ for estimating the
parameter of the correlation function, $\theta$.  So this MSPE criteria
balances reducing predictive variance with learning local wigglyness of the
surface.

It turns out that the contribution of the second term, beyond the new
reduced variance, is small. \citeauthor{gramacy:apley:2014} show that the full MSPE
criteria leads to qualitatively similar local designs $X_n(x)$ as ones
obtained using just $V_j(x|x_{j+1}; \hat{\theta}_j)$, which provides
indistinguishable out-of-sample predictive performance at a fraction of the
computational cost (since no derivative calculations are necessary).  This
simplified criteria is equivalent to choosing $x_{j+1}$ to maximize 
{\em reduction} in variance:
\begin{align}
 v_j(x&; \theta)  - v_{j+1}(x; \theta) \label{eq:dxy} \\
&= k_j^\top(x) G_j(x_{j+1}) m_j^{-1}(x_{j+1}) k_j(x) + 2k_j^\top(x)
  g_j(x_{j+1}) K(x_{j+1},x) + K(x_{j+1},x)^2 m_j(x_{j+1}), \nonumber
\end{align}
where $G_j(x') \equiv g_j(x') g_j^\top(x')$,
\begin{equation}
 g_j(x') = -m_j(x') K_j^{-1}
k_j(x') \;\;\;\;\; \mbox{and} \;\;\;\;\; m_j^{-1}(x') = K_j(x',x') -
k_j^\top(x') K_j^{-1} k_j(x'). \label{eq:piedef}
\end{equation}
Those $O(j^2)$ calculations are a simple consequence of deploying the
partitioned inverse equations on the salient elements of Eq.~(\ref{eq:newv}),
thereby bypassing more expensive $O(j^3)$ ones.  Although known for some time
in other contexts,
\citeauthor{gramacy:apley:2014} chose the acronym ALC to denote the use
of that decomposition in local design in order to recognize its first use towards {\em
 global} design of computer experiments by a method called {\em active
 learning Cohn} (\citeyear{cohn:1996}).  That scheme required numerically
 integrating ({\ref{eq:piedef}) over the entire design space. Although the
 localized analog above is simpler because it does not involve an integral,
 both global and local versions require a computationally intensive search 
 over a large set of candidates $x_{j+1} \in X_N \setminus X_j(x)$.  Speeding up this search is the primary focus of our
contribution.

Global emulation, that is predicting over a dense grid of $x$-values, can be
done in serial by looping over the $x$'s, or in parallel since each
calculation of local $X_n(x)$'s is independent of the others.  This kind of
embarrassingly parallel calculation is most easily implemented on symmetric
multiprocessor (SMP) machines via {\tt OpenMP} pragmas.
%, which make it trivial
%to execute each element of a {\tt for} loop with a new thread.  
As we
demonstrate in Section \ref{sec:big}, one can additionally divvy up predictions
on multiple nodes of a cluster for very big calculations.  Finally,
\citeauthor{gramacy:apley:2014} recommend a two-stage scheme wherein local
$\hat{\theta}_n(x)$'s are calculated after each local sequential design
$X_n(x)$ is chosen, so that the second iteration's local designs use locally
estimated parameters. This leads to a globally non-stationary model which
provides highly accurate predictions under a tight computational budget.
\begin{figure}
\centering
\fbox{
\begin{minipage}{16cm}
\begin{enumerate}
  \item Choose a sensible starting global $\theta_x = \theta_0$ for all $x$.
\item Calculate local designs $X_n(x, \theta_x)$ based on ALC,
  independently for each $x$:
\begin{enumerate}
\item Choose a NN design $X_{n_0}(x)$ of size $n_0$.
\item For $j=n_0, \dots, n-1$, set
\[
  x_{j+1} = \mathrm{arg}\max_{x_{j+1} \in X_N \setminus X_j(x)} v_j(x; \theta_x) 
  - v_{j+1}(x; \theta_x),
\]
and then update $D_{j+1}(x, \theta_x) = D_j(x, \theta_x) \cup (x_{j+1}, y(x_{j+1}))$.
\end{enumerate}
\item Also independently, calculate the MLE $\hat{\theta}_n(x) |
  D_n(x, \theta_x)$ thereby explicitly obtaining a globally nonstationary
  predictive surface. Set $\theta_x = \hat{\theta}_n(x)$.   
\item Repeat steps 2--3 as desired. 
\item Output predictions $Y(x)|D_n(x, \theta_x)$ for each $x$.
\end{enumerate}
\end{minipage}
}
\caption{Multi-stage approximate local GP modeling algorithm.}
\label{f:alg}
\end{figure}
The full scheme is outlined algorithmically in Figure
\ref{f:alg}.
It is worth remarking that the scheme is completely deterministic, calculating
the same local designs for prediction at $x$, given identical inputs ($n$,
initial $\theta_0$ and data $D_N$) in repeated executions.  It also provides
local uncertainty estimates---a hallmark of any approximation---via
Eq.~(\ref{eq:preds2}) with $D_n(x)$, which are organically inflated relative
to their full data ($D_N$) counterparts.
Empirically, those uncertainty estimates
over cover, as they are perhaps overly conservative.  
\citeauthor{gramacy:apley:2014} suggest adjustments
that can be made to project towards narrower bounds which are closer to their
full $N$ counterparts.

\section{GPU computing}
\label{sec:gpu}

Under NVIDIA's {\tt CUDA} programming model, work is offloaded to a general
purpose GPU device by calling a \emph{kernel} function---specially written
code that targets execution on many hundreds of GPU cores. {\tt CUDA} has
gained wide-spread adoption since its introduction in 2007 and many
``drop-in'' libraries for GPU-acceleration have been published, e.g., the
{\tt CUBLAS} library which contains a {\tt cublasDgemm} function that is the
GPU equivalent of the {\tt DGEMM} matrix-matrix multiplication function from
the {\sf C} {\tt BLAS} library. Such GPU-aware libraries allow for significant
speedups at minimal coding investment, and most use of GPUs for acceleration
in statistical applications has been accomplished by replacing calls to
CPU-based library functions with the corresponding GPU kernel call from a
GPU-aware library
\citep{franey:ranjan:chipman:2012,eidsvik2013estimation,paciorek:etal:2013}.
This can be an effective approach to GPU-acceleration when the
bottleneck in the program lies in manipulating very large matrices, e.g., of
dimension $\geq 1000$, as otherwise GPU-aware math libraries can actually be
less efficient than CPU ones. In our application, the calculations in
Figure
\ref{f:alg} involve relatively small matrices by design and therefore do
not benefit from this drop-in style approach to GPU-acceleration.
Instead, we have developed a custom kernel that 
%% implements the entirety of Step 2(b) in Figure \ref{f:alg} that 
is optimized for our relatively
small matrices and also carries out many processing steps in a single
invocation. 

The nuances of our implementation require an understanding of the
GPU architecture.  In the {\tt CUDA} computing model, threads are grouped into
\emph{blocks} of up to 1024 threads per block.\footnote{All values reported here 
are for CUDA Compute Capability version 2.0 which is the version used in our experiments.}
Up to $65535$ thread blocks can be instantiated to create the kernel 
\emph{grid}, a structure of
thread blocks on which a GPU kernel function is invoked.\footnote{We restrict
ourselves to a 1-d grid; more blocks may be instantiated in 2-3d grids.}
Groups of threads belonging to a given block are simultaneously
executed in a \emph{warp}.  All warps derived from a given block
are guaranteed to be resident on the same Streaming Multiprocessor (SM) on the
GPU device. The number of threads per warp is fixed by the GPU architecture
(our cards have 32 threads per warp) and the assignment of threads to warps is
controlled by the GPU hardware. The number of blocks that can run
simultaneously on a given SM is constrained by the amount of memory and the
number of registers required by the threads within each block. The number of
SMs and the total number of blocks is fixed by the GPU hardware architecture
(our cards have 16 SMs, and each can host multiple blocks simultaneously).
Assigning multiple blocks to a single SM allows threads from one block to
utilize the SM, e.g., perform floating point operations, while threads from
another block wait for memory transactions to complete.

Relative to other parallel architectures, GPUs allocate a relatively small
amount of memory and registers to each thread. In descending order of access
speed, the types of memory utilized for our kernels are registers, shared
memory, and local/global memory. Registers and local memory are
thread-specific and up to 32768 registers are available to the threads
belonging to a given block.  Shared-memory (up to 48KB per block) is
accessible by all threads belonging to the same block and provides a
high-speed location for threads within the same block to communicate with one
another and work collectively on data manipulation. Global memory (up to 5GB
per GPU device) is accessible by all threads across all blocks, but is an
order of magnitude slower than shared-memory and registers.  Because all
inter-block communication must use global memory, GPU-based applications tend
to only achieve high performance on strongly data-parallel algorithms in which
work can be cleanly divided across the thread blocks, thereby allowing them to
operate independently. For detailed information about parallelism and memory
in GPUs, please see \cite{kirk2010programming}.

Due to the multiple levels of parallelism, and the different memory types and
speeds, constructing kernels can be difficult and, sometimes, counterintuitive.
In the remainder of this section, we isolate the calculations from Figure
\ref{f:alg} that are best suited to the GPU architecture, describe how
those can be implemented on a GPU, and discuss how best to
utilize the resulting GPU subroutine in the wider context of global
approximate emulation.

\subsection{GPU ALC calculation}
\label{sec:alcgpu}

The most computationally intensive subroutine in the local approximate GP
algorithm is Step 2(b) in Figure \ref{f:alg}: looping over all remaining
candidates and evaluating the reduction in variance (\ref{eq:dxy}) to find the
next candidate to add into the design.  Each reduction in variance calculation
is $O(j^2)$, and in a design with $N$ points, there are $N' = N-j$ candidates.
Usually $N \gg j$, so the overall scheme for a single $x$ is $O(Nn^3)$, a
potentially huge undertaking called for $j=n_0, \dots, n$ for each predictive
location $x$. As \cite{gramacy:apley:2014} point out, it may not be necessary
to search over all $N-j$ candidates---searching over a smaller set (say $N' =
100n$) of NNs can consistently yield the same local design as searching over the
full set. However, the resulting $O(n^4)$ search can still represent a
considerable computational undertaking, even for modest $n$, when the number
of predictive locations is large.

The structure of the evaluations of \eqref{eq:dxy}, independent for each of
the $N'$ candidates, is ideal for GPU computing.  Each candidate's calculation
can be assigned to a dedicated thread block so long as $N' < 65535$, i.e. the number of thread blocks. Each of
the sequence of $O(j^2)$ operations required for each candidate's calculation
(i.e., each block) can be further parallelized across $j$ threads within the
designated block so long as $j \leq n < 1024$, potentially in parallel with
many others. Some care is needed to ensure that (a) as many of these
independent calculations as possible actually {\em do} occur in parallel; (b)
threads execute the same instructions on nearby memory locations at the same
time for high throughput; (c) there are as few synchronization points as
possible; (d) memory transfers to and from the GPU device are minimized; and
(e) memory accesses on the GPU are primarily to fast memory locations rather
than to high-latency global memory.

\begin{figure}[ht!]
\centering
\fbox{
\begin{minipage}{16cm}
The thread is indexed by $t$, and the block by $b$.\\

Scalar inputs stored in {\bf registers}:
\begin{center}
\begin{tabular}{r|l}
variable & description \\
\hline
$j$ & integer number of rows in the current local design $X_j(x)$ \\
$\theta$ & double precision lengthscale parameter $\theta$ \\
$\eta$ & double precision nugget parameter $\eta$ \\
$N$ & the number of rows $(N-j)$ in the candidate matrix $\tilde{X} = X_N \setminus X_j(x)$ \\
$p$ & integer number of columns in $X_j(x)$ and $\tilde{X}$ \\
\end{tabular}
\end{center}

\medskip
Double-precision input (and output) arrays stored in {\bf global memory}:
\begin{center}
\begin{tabular}{r|l}
variable & description \\
\hline
$X$ & row-wise flattened $X_j(x)$, a $j \times p$ matrix \\
$K^{-1}$ & row-wise flattened  $K^{-1}$, a $j \times j$ matrix \\
$\tilde{X}$ & row-wise flattened $X_N \setminus X_j(x)$, a $(N-j) \times p$ matrix \\
$h$ & covariances $K(x, X)$ between $x$ and rows of $X$, an $n$-vector\\
$\Delta$ & an $N-j$ vector containing the output of Eq.~(\ref{eq:dxy})
\end{tabular}
\end{center}

\medskip
Double-precision working memory scalars stored in {\bf registers}:
\begin{center}
\begin{tabular}{r|l}
variable & eventual contents via analog in Eq.~(\ref{eq:dxy}) \\
\hline
$m^{-1}$ & $m_j^{-1}$, identical for all threads %$t$ 
in block $b$ \\
$k_t$ & $(k_j(x))_t$, the $t^\mathrm{th}$ element of $k_j(x)$ \\
& later re-used for the $t^\mathrm{th}$ entry of $K_j^{-1} k_j(x_b)$\\
$g_t$ & $(g_j(x_b))_t$, the $t^\mathrm{th}$ element of $g_j(x_b)$ \\
$\ell_t$ & $K(x_b, x)$, identical for all threads $t$ in block $k$ \\
& later re-used for the $t^\mathrm{th}$ entry of  $G_j^{-1}(x')m_j^{-1}(x_b) k_j(x)$
\end{tabular}
\end{center}

%\medskip
%Scalar integer quantities are used for indexing but they are not listed here.

\medskip
Double-precision working memory arrays stored in {\bf shared memory}:
\begin{center}
\begin{tabular}{r|l}
variable & eventual contents via analog in Eq.~(\ref{eq:dxy}) \\
\hline
$x_b$ & a $p$-vector: the $b^\mathrm{th}$ candidate/row of $\tilde{X}$\\
$k$ & $k_j(x)$, a $j$-vector; \\
& later re-used for element-wise product of $k_j(x_b)$ and $K_j^{-1} k_j(x_b)$\\
$g$ & $g_j(x_b)$, a $j$-vector \\
$\ell$ & a $j$-vector with  element-wise product of $k_j(x)^\top$ and $K_j^{-1} k_j(x_b)$; \\
& later re-used for product of $k_j(x)^\top$ and $G_j^{-1}(x')m_j^{-1}(x_b) k_j(x)$
\end{tabular}
\end{center}
\end{minipage}
}
\caption{Inputs, outputs and working memory used by the GPU kernel computing
Eq.~(\ref{eq:dxy}).}
\label{f:io}
\end{figure}

\begin{figure}[ht!]
\centering
\fbox{
\begin{minipage}{16cm}
Recall that $t$ indexes the thread and $b$ indexes the block.\\

Each enumerated set of instructions is implicitly followed by a thread synchronization.
\begin{enumerate}
\item {\em \% Copy the $b^\mathrm{th}$ candidate (row of $\tilde{X}$) into faster shared memory.}\\
if $(t < p)$ then $x_b[t] \leftarrow \tilde{X}_b[b\times p + t]$

\item {\em \% Calculate $K_j(x_b, x)$.}\\
$k_t \leftarrow 0$\\
for $(i \in \{1:p\})$ do $k_t \leftarrow k_t + (x_b[i] - X[t\times p+i])^2$\\
$k[t] \leftarrow \exp\{-k_t/\theta\}$

\item {\em \% Initialize $g_j(x_b)$ with $K_j^{-1}k_j(x_b)$, and
prepare $k_j(x_b)^\top K_j^{-1}k_j(x_b)$.}\\
$g_t \leftarrow 0$\\
for $(i \in \{1:j\})$ do $g_t \leftarrow g_t + k[i] \times K^{-1}[i\times j + t]$.\\
$\ell[t] \leftarrow g_t \times k[t]$

\item {\em \% Complete the dot product $k_j(x_b) \cdot K_j^{-1}k_j(x_b)$.}\\
$\ell[0] \leftarrow \mathrm{sum.reduce}(t, \ell)$

\item {\em \% Calculate $\mu^{-1}_j(x_b)$, and finish $g_j(x_b)$.}\\
$m^{-1} \leftarrow 1.0 + \eta - \ell[0]$\\
$g[t] \leftarrow g_t/m^{-1}$

{\em \% Without syncing threads, calculate $K_j(x, x_b)$ and initialize the output array.}$^*$ \\
$k_t \leftarrow 0$\\
for $(i \in \{1:p\})$ do $k_t \leftarrow k_t + (x_b[i] - x[i])^2$ \\
$\Delta[b] \leftarrow \exp \{ -k_t/\theta \}$ 

\item {\em \% Prepare $k_j(x)^\top G_j^{-1}(x_b) k_j(x)$ and $k_j^\top(x) g_j(x_b)$.}\\
$\ell_t \leftarrow 0$
for $(i \in \{1:j\})$ do $\ell_t \leftarrow \ell_t + g[t] \times g[i] \times \mu^{-1}_t$\\
$\ell[t] \leftarrow \ell_t \times h[t]$\\
$k[t] \leftarrow h[t] \times g[t]$

\item {\em \% Complete the dot products $k_j(x) \cdot G_j^{-1}(x')m_j^{-1}(x_b) k_j(x)$ and 
$k_j(x) \cdot K_j^{-1} k_j(x_b)$.} \\
$\ell[0] \leftarrow \mathrm{sum.reduce}(t, \ell)$\\
$k[0] \leftarrow \mathrm{sum.reduce}(t, k)$

\item {\em \% Wrapping up Eq.~(\ref{eq:dxy}).}$^*$\\
$\Delta[b] \leftarrow \ell[0] + 2.0\times k[0] \times \Delta[b] + \Delta[b]^2/\mu^{-1}$

\end{enumerate}
\end{minipage}
}
\caption{Psuedocode for the GPU kernel computing
Eq.~(\ref{eq:dxy}).  Comments are indicated by lines beginning with
``\%''; those followed by ``*'' superscripts indicate that the following
commands need only be executed on one thread, e.g., thread $t=0$. 
\label{f:gpualg}}
\end{figure}

Figures \ref{f:io} and \ref{f:gpualg} describe our GPU implementation via the
data/memory structure and the execution sequence, respectively.  In both cases
the description is for a particular block and thread within the block, indexed
by $b$ and $t$ respectively.  The $b$ index selects a candidate $x_b$ from a
row of the set of remaining candidates $\tilde{X}_j(x) = X_N\setminus X_j(x)$.
When the kernel executes, many blocks, i.e., a range of $b$-values, are run in
parallel. The number which execute in parallel depends on the size of the
problem, $j$ and $N$, and other operating conditions, but we find that it is
typically in the hundreds for the problems we've attempted. Within a block,
the $t$ index selects a column of a matrix, or an entry of a vector, in order
to parallelize the within-block computation.  Based on the value of $t$,
threads can take different execution paths.  However,
execution is swiftest when $\approx 32$ threads (the warp size) execute the
same sequence of operations on adjacent memory locations.  Therefore an effort
has been made to avoid divergent execution paths whenever possible.

Figure \ref{f:io} describes the inputs/outputs (first two tables) and the
working memory (last two) of the GPU kernel.  The right-hand columns of the
tables describe variables in terms of quantities in
Eq.~(\ref{eq:dxy}). Note that some are reused.  It also indicates
what type of memory the variable is stored in.  Initially, all non-scalar
inputs reside in slow global memory.  Parts of global memory that are
frequently accessed by the block, $b$, are copied into that block's shared
memory. Shared memory locations which are repeatedly accessed by particular
threads, $t$, within a block use temporary register storage.  No local memory
is required for our kernels.  As a visual queue we use a $t$ subscript to
distinguish between a register quantity indexing a particular value of a
shared memory array.  For example, $k_t$ is used to calculate what will
eventually reside in $k[t]$, the $t^\mathrm{th}$ indexed shared memory mapping
pointed to by $k$. Eventually, $k$ will store $k_j(x_b)$, a $j-$vector, and
will later be reclaimed to store $K_n^{-1}k_j(x_b)$.

Several of the steps outlined in Figure \ref{f:gpualg} require more detailed
explanation.  Notice that Steps 2 \& 6 assume an isotropic Gaussian
correlation function.  Simple modification would accommodate another family
and/or a separable version via a vectorized $\theta$ parameter.  In two
places, a sequence of two synchronized steps ($3 \rightarrow 4$ and $6
\rightarrow 7$) calculate the scalar value(s) of a quadratic form by first
having each thread, asynchronously, fill a particular entry in a $j$-vector,
and then sum its elements via a {\em reduction}. Reductions are a way to
get multiple threads to work simultaneously towards calculating something that
is more natural serially, like a sum.  Our implementation, abstracted as
``sum.reduce($t$, $v$)'' for thread $t$'s contribution to calculating the
sum of the vector $v$, uses the logarithmic
version described on the \citet{sharcnet:cuda} pages, which makes use of
$\lfloor j/2 \rfloor$ threads.  Since more than half of the threads are idle
in this reduction, we implemented our own bespoke version (employing the idle
threads) for the two simultaneous reductions required by Step 7, which led to
a 10\% speedup compared to two separate reductions.  [See Appendix
\ref{sec:sum2}.]

Step 3 is the most computationally intensive, since it involves accessing $j$
items stored in global memory, the $t^\mathrm{th}$ column of $K^{-1}$. There
is one other $j$-loop (Step 6), but it accesses faster shared memory.  Staging
the $t^\mathrm{th}$ column of $K_j^{-1}$ in shared memory does not lead
to a faster implementation since multiple accesses of this data are not required within
the block.  By contrast, we copy a row of $\tilde{X}$ in to
shared memory (Step 1) since it is reused (Step 5), thereby avoiding multiple 
transactions on global memory in this phase of the algorithm.  We remark that it is very
important to work column-wise with $K^{-1}$ as opposed to row-wise to ensure coalescence in memory transactions.  Working column-wise 
allows warps of threads to access adjacent memory locations storing $K^{-1}$.
Working row-wise, i.e., accessing $K[t\times j + i]$, gives the same answer
(because the matrix is symmetric) but is about $j$-times slower.

Finally, we remark that the output, $\Delta$, is not normalized.  A final
step, multiplying by $\psi/(j-2)$, is required to match the expression in
Eq.~(\ref{eq:dxy}).  This can be done as a CPU post-processing step, although
it is slightly faster on the GPU.  In Figure \ref{f:gpualg}:
\begin{enumerate} \singlespacing
\item[9.] {\em \% Normalize by the global variance estimate.}\\
$\Delta[b] \leftarrow \psi \Delta / (j-2)$.
\end{enumerate}
Observe that this is not actually required to find the $\mathrm{argmax}$ in
Step 2b of Figure \ref{f:alg}.

\subsection{GPU--CPU full GP approximation}

The GPU kernel described above implements Steps 2a and 2b in Figure
\ref{f:alg}.  Here we discuss how it can be best situated within the outer
loop(s), ultimately being applied over all predictive locations $x \in
\mathcal{X}$.  The simplest option is to serialize: simply calculate for each
$x$ in sequence, one after another. Within that loop, iterate over $j=n_0,
\dots, n$, performing the required CPU calculations amidst GPU kernel calls to
calculate Eq.~(\ref{eq:dxy}). We show in Section \ref{sec:results} that this
leads to significant speedups compared to a serial CPU-only implementation.
But it makes for an inefficient use of a multiplicity (i.e., 1-CPU and 1-GPU)
of computing resources.  The CPU is idle while the GPU is working, and vice
versa.  

% I'm not sure this bit about async memory copies is true.  You can 
% do async memory copies to-from the device with cudaMemCpyAsync()
% and this is a common trick to avoid copy overhead, but i don't recall
% talking with you about implementing this.  Did you use cudaMemCpyAsync() in this code?
% if not, cudaMemCpy() is blocking on the CPU.  Kernel launches are async tho.

Both inefficiencies are addressed by deploying a threaded CPU version
identical to the original one advocated by \cite{gramacy:apley:2014}, i.e.,
using {\tt OpenMP}.  The difference here is that speedups are attained even in
the case of a {\em single} CPU core because while one CPU thread is waiting for a
GPU kernel to finish other CPU threads can be performing CPU operations and/or
queuing up the next GPU calculations.  Having a small backlog of GPU kernels
waiting for execution is advantageous because it means the next kernel will start
immediately after the current one finishes.  
%% Moreover, expensive GPU memory
%% transfers can execute in parallel with other transfers {\em and} in tandem
%% with a running kernel.  Both reduce idling of the GPU multiprocessors.

There are diminishing returns for increasing numbers of CPU threads as they
compete for resources on both the CPU and GPU.  There would eventually be
negative returns due to inefficiencies on the CPU (too many context switches)
or GPU (not enough memory to queue executions).  This latter concern is very
unlikely though since, e.g., our device has more than 5GB of global memory.
When $N' < 65535$ and $j \leq n < 1024$, i.e., the block and thread
constraints, we can still queue quite a few kernels.  In typical
approximations $N'$ and $n$ are an order of magnitude smaller and we find that
(marginal) speedups are still observed when there are more than 4 CPU threads
per CPU core (in the 1-GPU case).

Obviously, when there are multiple CPUs and/or multiple GPUs, CPU threading is
essential lest the duplicated resources remain untapped.  We assume here that
all GPUs are identical\footnote{Modern multi-core CPUs are always identical
when in the SMP configuration.} and, so long as the CPU threads spread the
kernels roughly equally amongst GPUs, no further load balancing considerations
are required \citep{hagan:2011}.  Given a fixed number of GPUs (including zero
for a CPU-only version) we find a nearly linear speedup as CPU cores (with one
thread each) are added.  Multiple threads-per-CPU core can help, although only
marginally as the number of cores increases, and only if there is at least one
GPU.  For example, we will show that 32 threads with 16 cores and one or two GPUs is
marginally faster than using 16 threads---one per core.  As GPUs
are added the initial benefits are substantial, especially when there are few
CPUs.  In that case it again makes sense to have more threads than CPU cores.

\section{Empirical results for the new GPU version}
\label{sec:results}

In this section we borrow the 2-d data and experimental apparatus of
\cite{gramacy:apley:2014}.  This allows us to concentrate on timing results
{\em only}---the accuracy, etc., of the resulting predictions are identical to
those reported in that paper.  Our discussion is broken into two parts: first
focusing on the ALC calculations in isolation; then as applied in sequence to
build up local designs for many input locations, independently (and in
parallel).  The node we used contains two NVIDIA Tesla M2090 GPU devices with
5GB of global memory and the L1 cache option is set to prefer shared-memory
(giving 48KB per block).  It has dual-socket 8-core 2.6 GHz Intel Sandy Bridge
Xeons with 32GB of main memory. 

\subsection{GPU calculations}
\label{sec:egpu}

Here we study the performance of  GPU ALC calculations [Section
\ref{sec:alcgpu}] relative to the one-CPU-only alternative, beginning with Figure \ref{f:alcgpu}
which summarizes the result of an experiment set up in the following way.  We
focus on a single reference location, $x$, the value of which is not important
as the timing results are the same for any $x$.   We consider $N'=60$K, which
is close to the maximum number of blocks, with one block per candidate.  The
only thing that varies is the local design size $n$, from $n=16$ to $n=512$
(in steps of size 4).  All of the required correlation matrices, etc., are
presumed to be calculated in advance (conditional on sub-designs $D_n(x)$, and
candidates $\tilde{X}$).  The time needed to build these is not included in
the comparison, as they are required as inputs by both CPU and GPU methods.
The extra time needed to copy data from CPU to GPU is, however, included in
the GPU timings.
  
\begin{figure}[ht!]
\centering
\includegraphics[scale=0.4,trim=0 0 30 0]{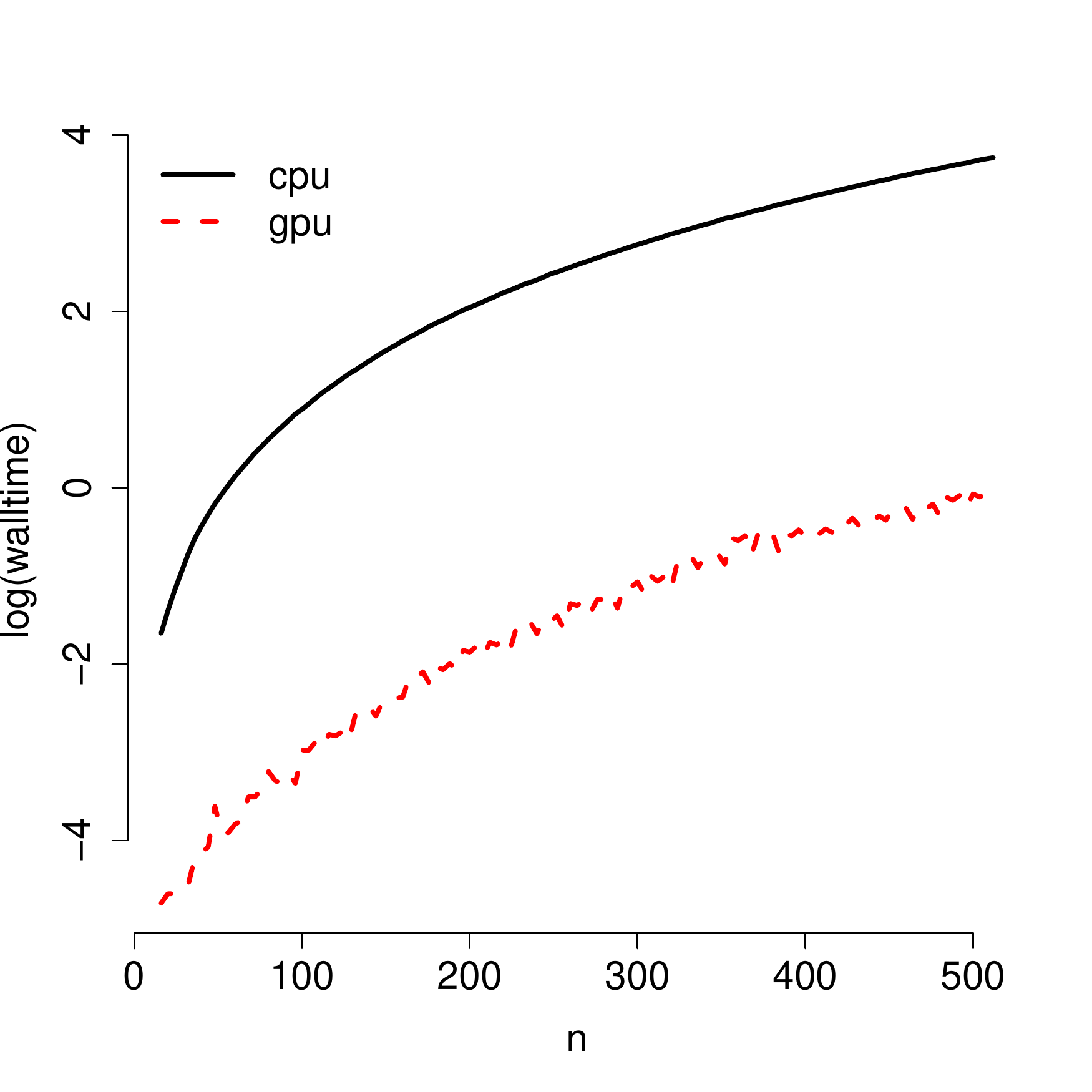} \hspace{1cm}
\includegraphics[scale=0.4,trim=28 0 30 0]{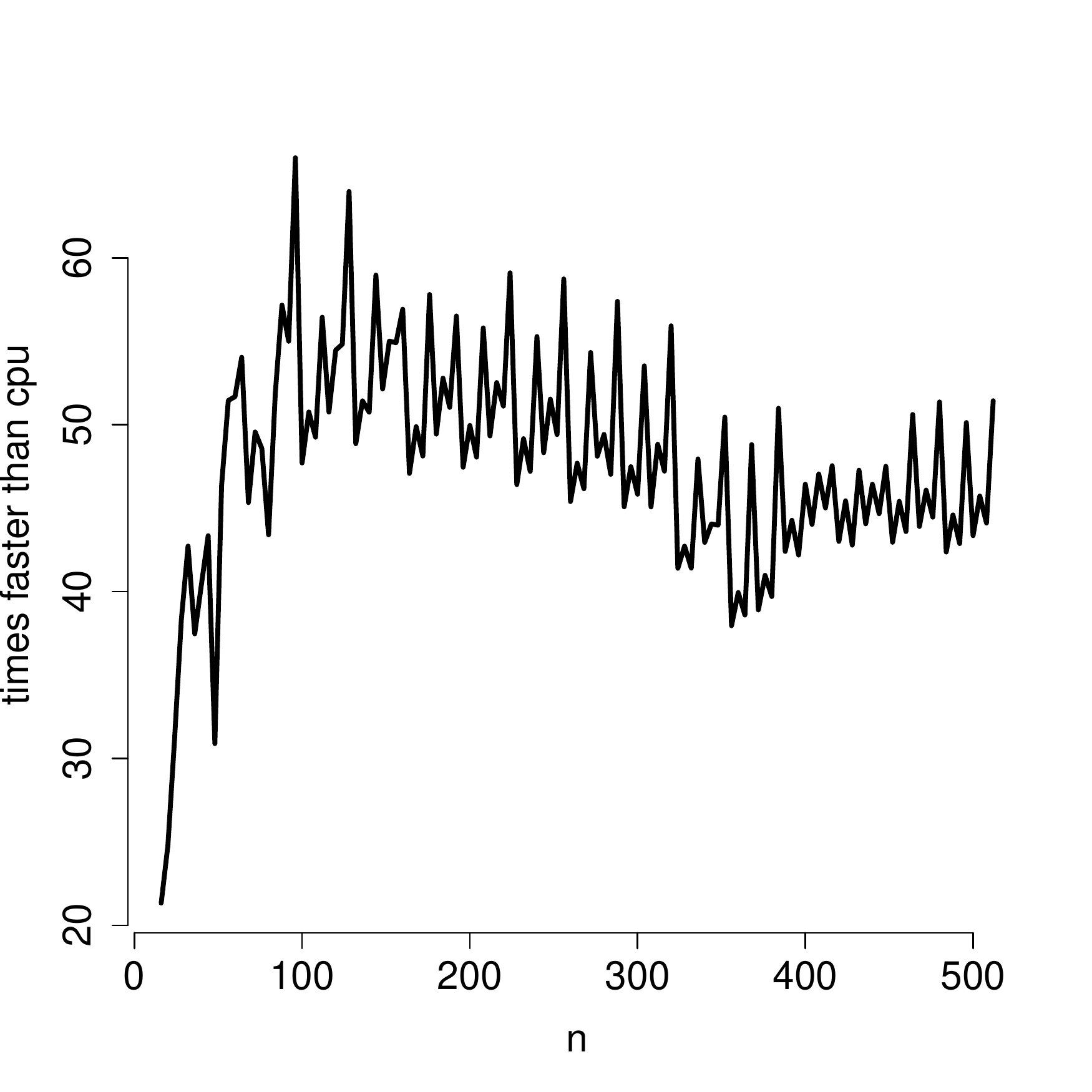} 
\caption{Comparing CPU-only and GPU-only wall-clock timings
for the ALC calculation at each of $n_{\mathrm{cand}}=60000$ candidate
locations for varying $n$, the size of the local design.  It is assumed that
the relevant covariance structure(s) are precomputed, so these are not
incorporated into the timing results.  The {\em left} plot shows absolute
performance on a $\log$ scale; the {\em right} plot makes a relative
comparison via ratios of times.}
\label{f:alcgpu}
\end{figure}

The {\em left} panel of the figure shows $\log$ timings separately for the CPU
and GPU; whereas the {\em right} panel shows the relative speedup offered by
the GPU obtained by dividing the CPU time by the corresponding GPU one.   We
observe the following. The speedups range between 20x and 75x, with more
modest speedups for small $n$ owing to fewer economies of scale. Two
explanations are: (a) initiating data transfers to the GPU are relatively
expensive when only a small amount of data is sent/received; and (b) when $n$
is small the number of GPU threads-per-block (also $n$) is low, which results
in low GPU throughput. We note that it may be possible to increase performance
in the second case, for low block index $b$,  by implementing more complex
thread allocation logic to increase the number of threads-per-block.  However,
we favor a direct mapping that leads to a clear implementation over a more
complex but potentially more performant solution. 

Observe [Figure \ref{f:alcgpu}, {\em right}] that the  time series of GPU
compute times is periodic, which is a consequence of the GPU architecture.
There are several factors contributing to this phenomena.  Executions where
$n$ is a multiple of the warp size, i.e., $n=32k$, tend to be faster, on
average. Our reduction scheme (for dot products, etc.) is fastest when $n$ is
a power of 2. There is a sweet spot near $n=128$, a power of 2 and multiple of
32, with diminishing returns thereafter due to the quadratically growing
$K^{-1}_n$ that must be transferred to the GPU over the relatively slow PCIe
bus.

% Finally, it is often the case that GPU kernels work best with 128 or 256
% threads due to the availability of registers per block. COMMENT: this a
% pretty hand-wavey statement.  It is safe to say the GPUs achieve high
% performance when there are enough threads to keep the hardware cranking along
% while other threads stall on memory transactions.  I think however we have
% provided sufficient explanation about why the timing curve looks the way it
% does at this point.

\subsection{Full GP approximations via CPU and GPU}
\label{sec:fullapprox}

Here we study how
the entire local design scheme, deploying extensive ALC search as a subroutine,
compares under GPU versus CPU when calculated for a dense set of predictive 
localtions $x$ in a global prediction exercise.  As in Section \ref{sec:egpu}, 
we primarly vary the approximation fidelity via the local design size, $n$.

 A quick profiling of a CPU-only version, varying $n$ and $x$, reveals that
ALC computations represent the majority of compute cycles, at 50--98\%.  That
wide range arises due the relative amount of ALC work required compared to
other calculations. For example, each iteration $j$ requires CPU routines to
update the GP correlation structure. At the end, when $j=n$, MLE calculations
may also be invoked.  Both can require a relatively substantial number of
cycles depending on the value of $j$, $n$, and the quality of initial
$\theta_x$ values, with bad ones leading to more likelihood evaluations to
find the MLE. As $j$, $n$, and $N'$ are increased, leading to a higher
fidelity approximation, the computational demands increase for ALC relative to
the other CPU routines, leading to more impressive speedups with a GPU
implementation.

\begin{figure}[ht!]
\centering
\includegraphics[scale=0.5,trim=10 0 0 20]{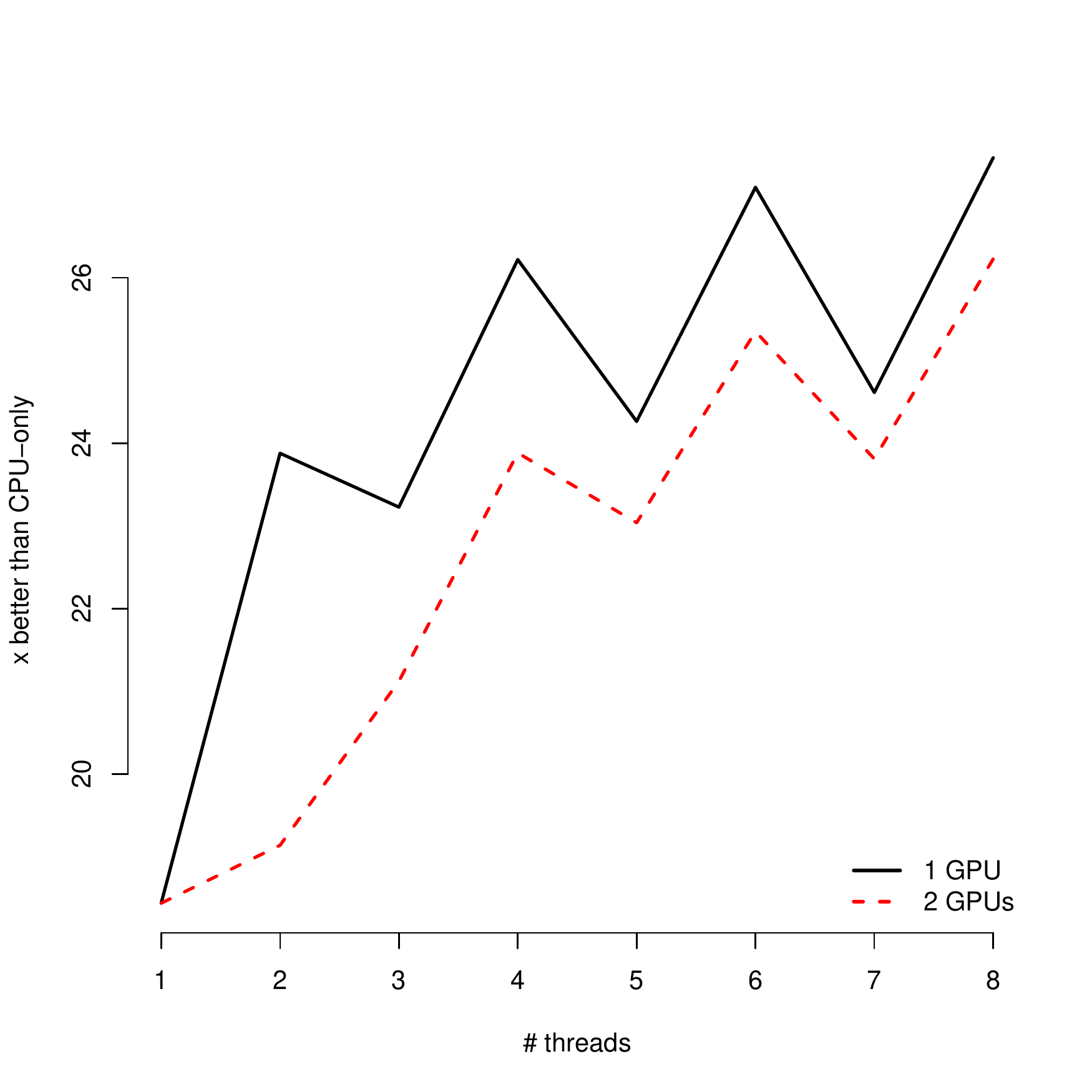}
\caption{Speedups from GPU version(s) for varying CPU threads on
a 1-core CPU.} 
\label{f:onecpu}
\end{figure}

To start things off, Figure \ref{f:onecpu} compares a CPU-only version to one
leveraging one- and two-GPU ALC calculations.  The metric shown on the
$y$-axis is a ratio of the wall-clock execution times obtained while predicting
at $\sim\!10$K locations, with $n=50$ and using $N' = 1000$ NN candidates.  The
numerators in that ratio are times calculated from the reference
implementation, a {\em single} CPU core without GPU(s).  The denominators come
from each CPU-GPU competitor, with one or two GPUs respectively, so that we
may interpret the ratio as a factor of improvement over a single-CPU-core-only
implementation. (Otherwise the setup is identical to the previous subsection.)
The $x$-axis varies the number of {\tt OpenMP} CPU threads, where each thread
works on a different local predictive location $x$, as described briefly in
Section
\ref{sec:laGP}. Threaded computing does not benefit a single-CPU-core-only
version. However, since GPUs can execute in parallel with other CPU
calculations, the entire scheme benefits from having multiple CPU threads
because they can asynchronously queue jobs on the GPU devices.  The trend is
that speeds increase, with diminishing returns, when the number of threads is
increased. Notice how odd numbers of threads are sub-optimal, which is a
peculiarity of the particular GPU-CPU architecture on our machine. Also,
notice how the 2-GPU setup is marginally slower than the 1-GPU one, suggesting
that the single CPU core is not able to make efficient use of the second GPU
device.

The previous example illustrates how CPU-only and CPU/GPU schemes might
compare on an older-model laptop or desktop connected to a modern GPU.  These
days even laptops are multi-cored (usually two cores), and most modern
research workstations have at least four cores.  It is not uncommon for them
to have up to sixteen. Therefore we next factor SMP-style parallelization into
the study.  In some ways this leads to a fairer CPU vs.~GPU comparison,
since GPUs are technically multi-core devices (albeit with very simple cores).
\begin{figure}[ht!]
\centering
\includegraphics[scale=0.336,trim=0 35 30 0,clip=TRUE]{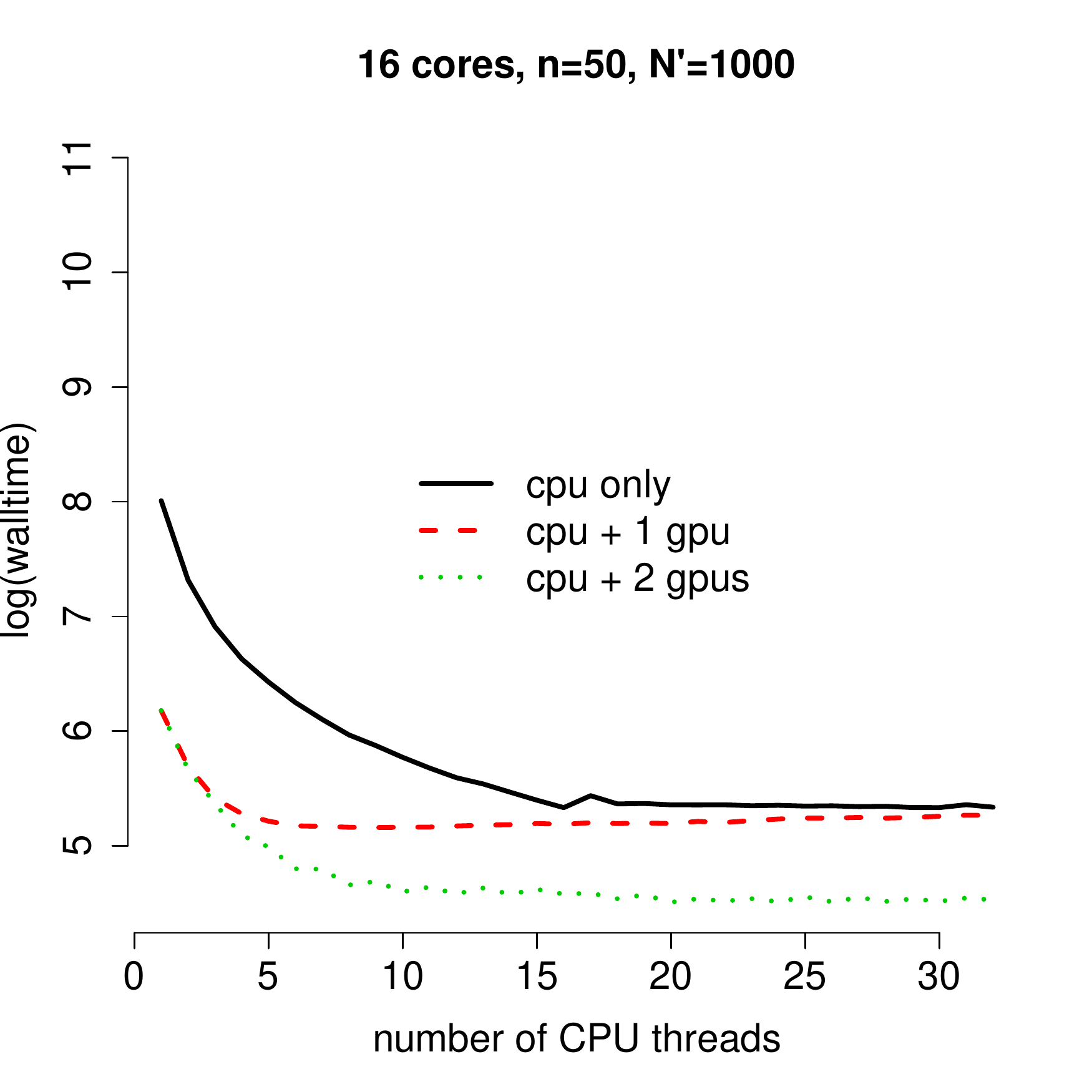} \hfill
\includegraphics[scale=0.336,trim=28 35 30 0,clip=TRUE]{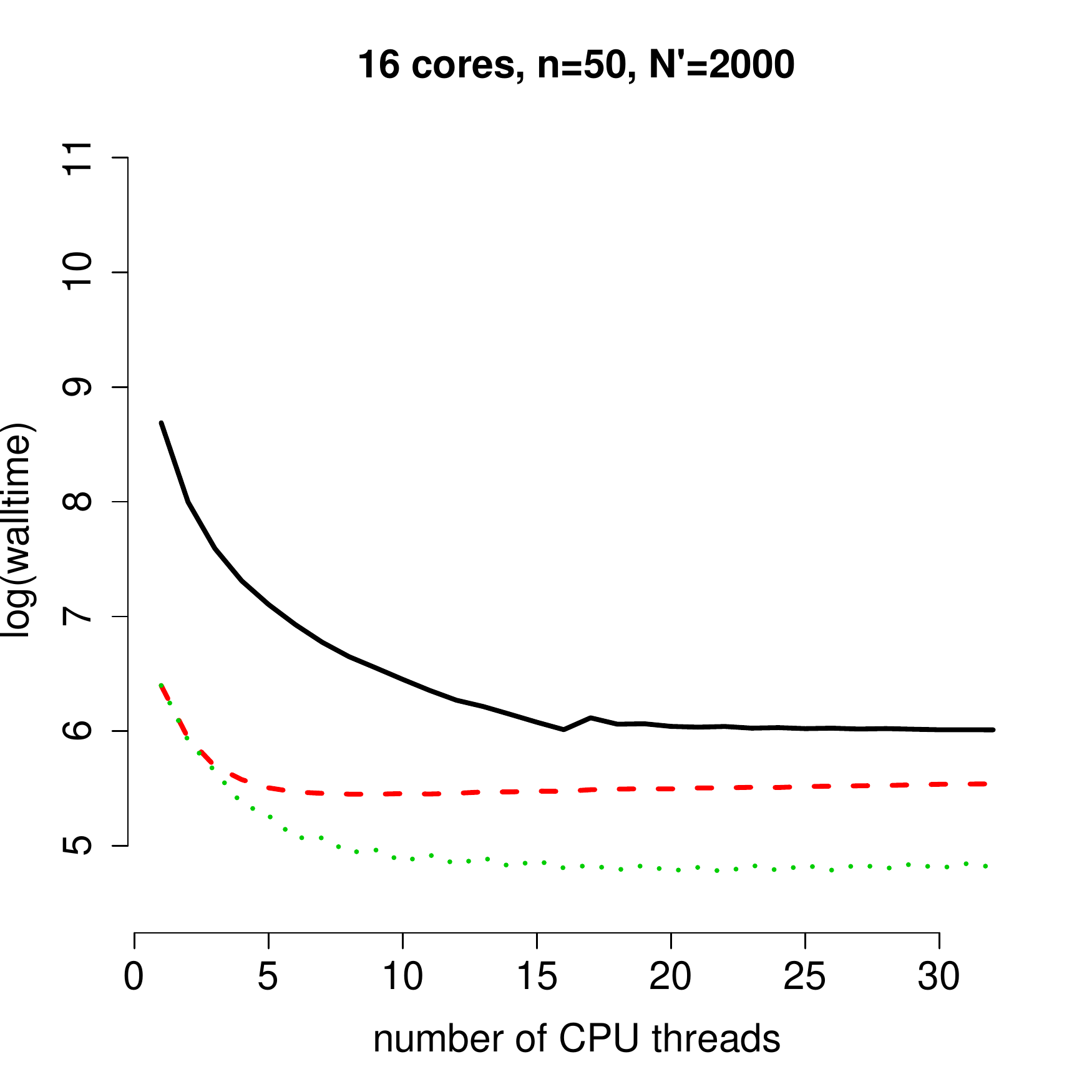} \hfill
\includegraphics[scale=0.336,trim=28 35 30 0,clip=TRUE]{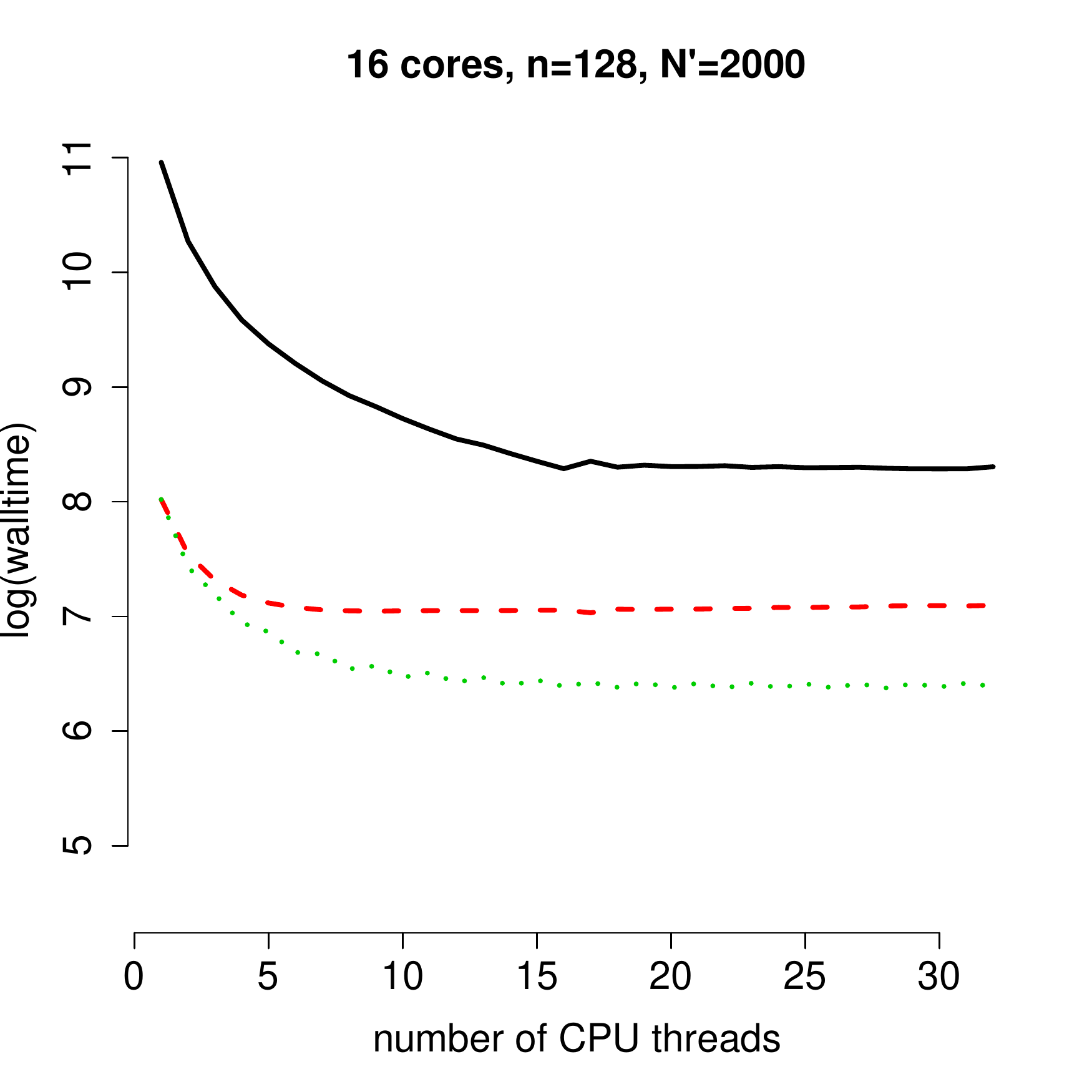}

\includegraphics[scale=0.336,trim=0 0 30 40,clip=TRUE]{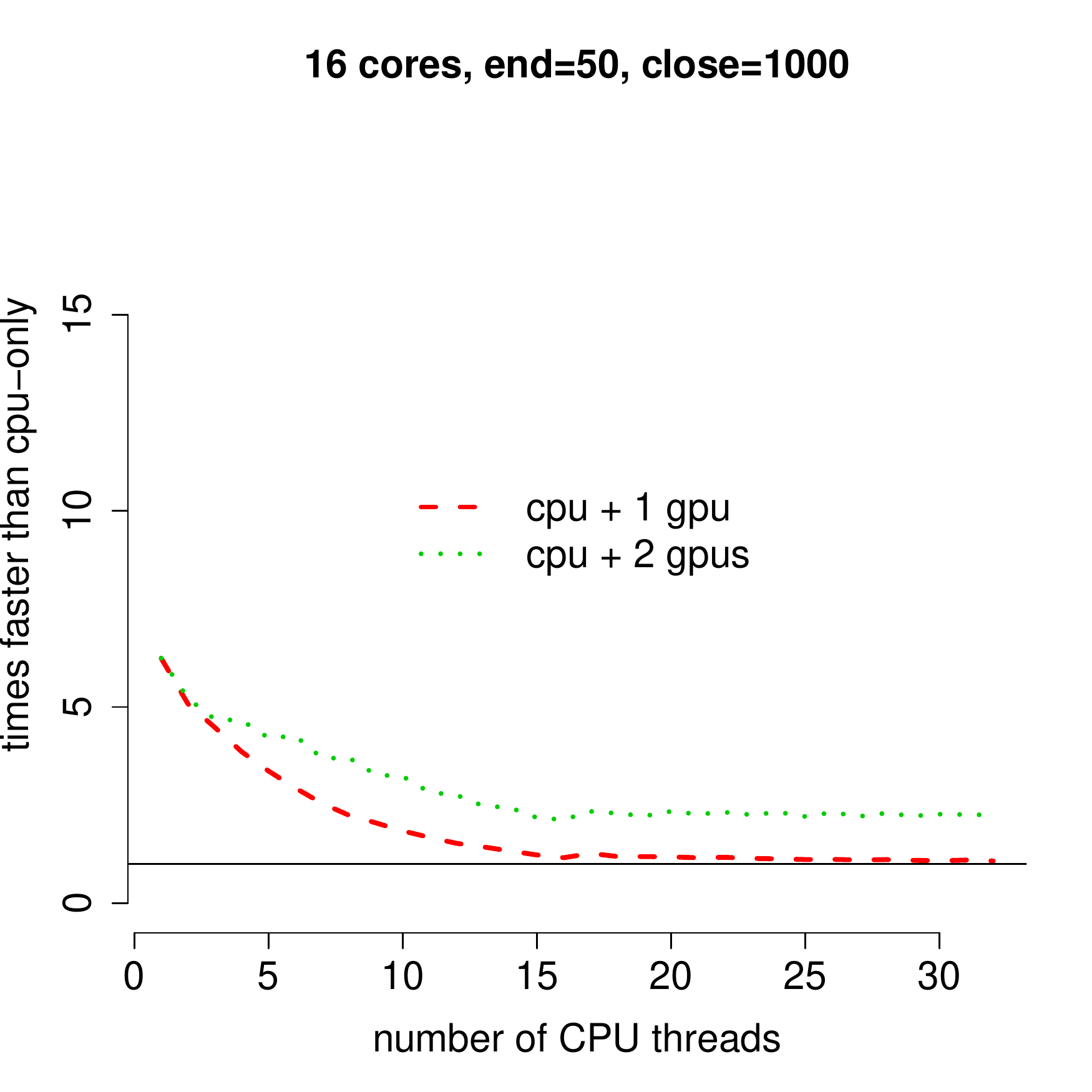} \hfill
\includegraphics[scale=0.336,trim=28 0 30 40,clip=TRUE]{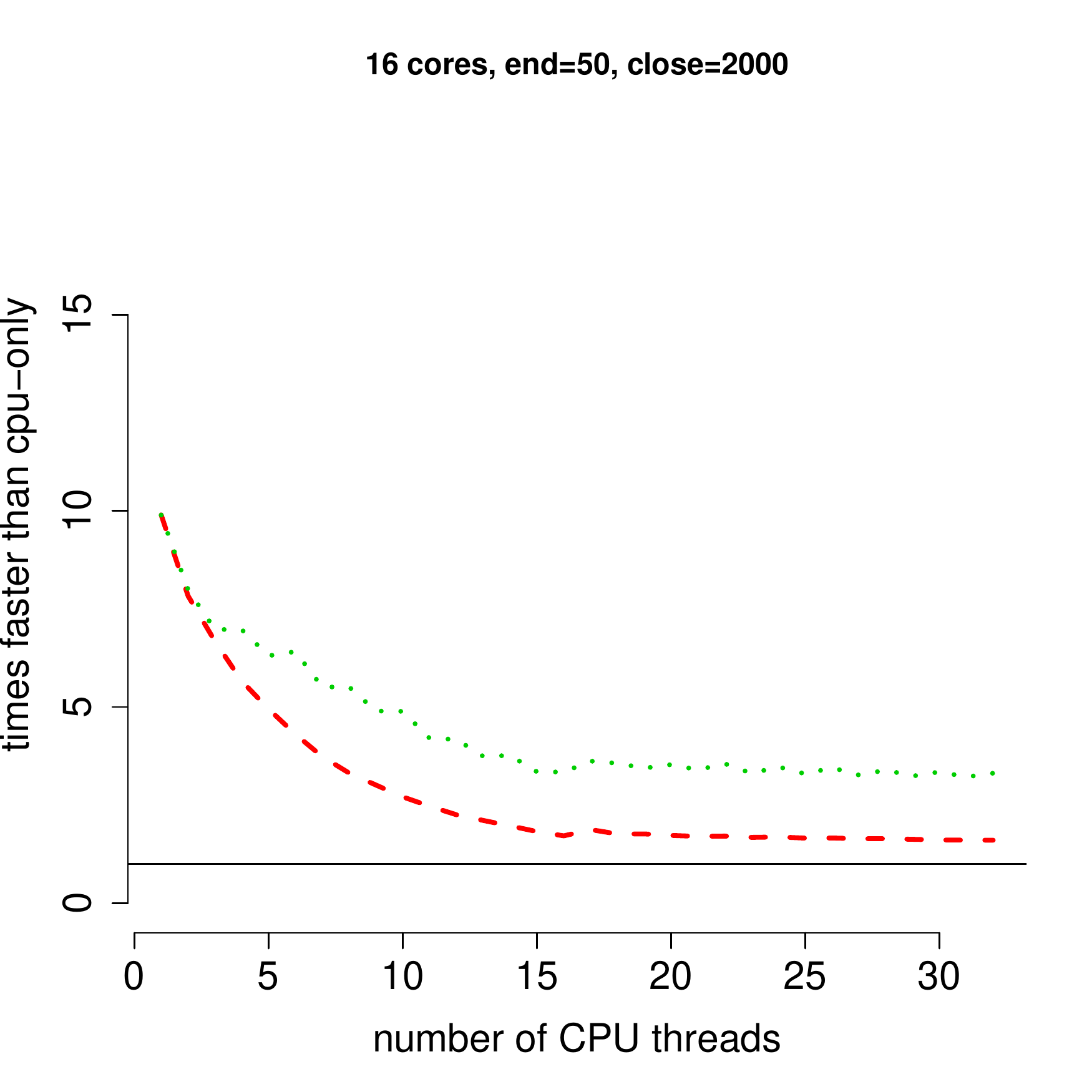} \hfill
\includegraphics[scale=0.336,trim=28 0 30 40,clip=TRUE]{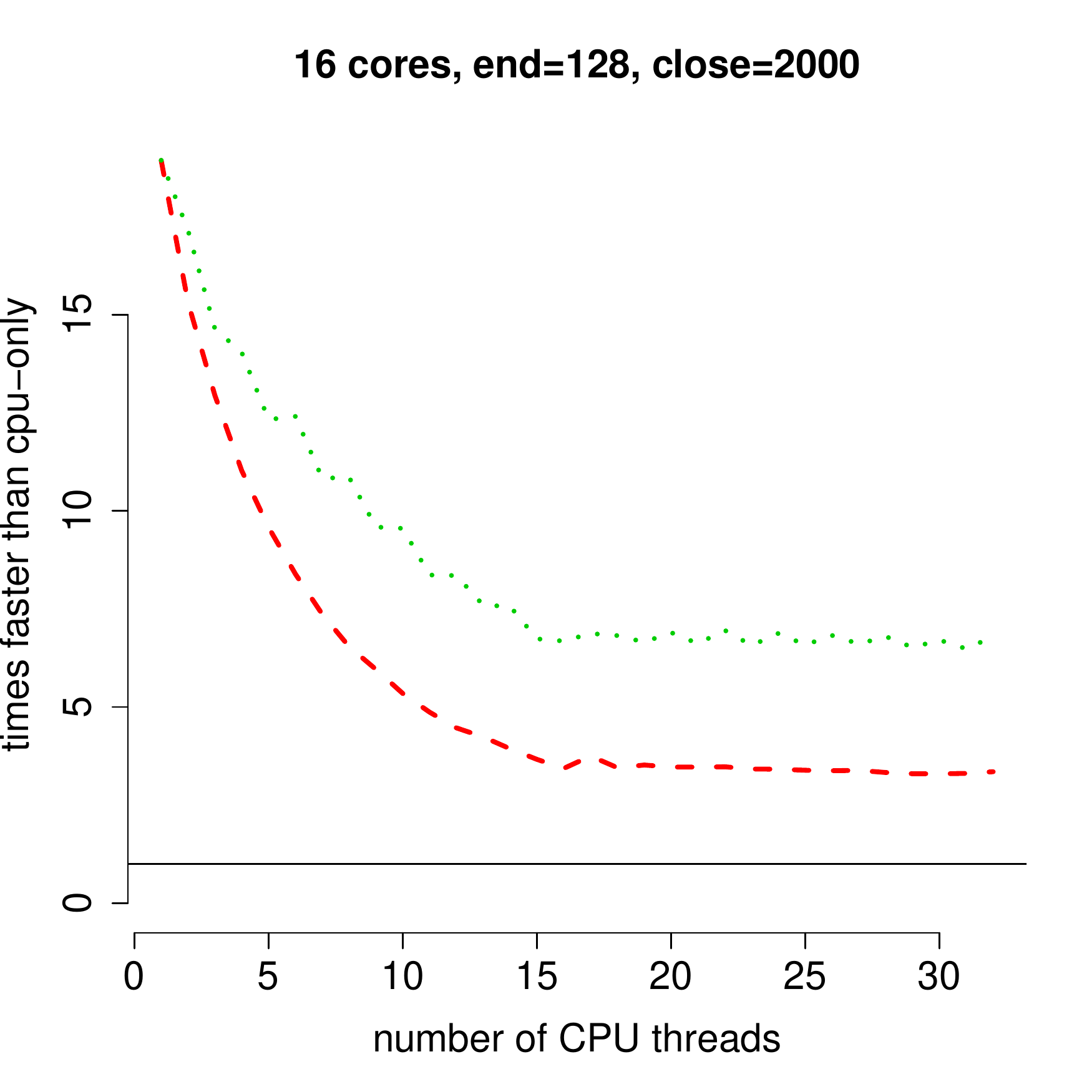}
\caption{Comparing full global approximation times for a $\sim40$K-sized
design, and a $\sim 10$K predictive locations, under various CPU/GPU
configurations and approximation fidelities (larger $n$ and/or $N'$).  The
{\em top} row makes absolute comparisons via $\log$ compute times; the {\em
bottom} row makes a relative comparison to CPU-only compute times via a
ratio. The solid horizontal line is at one on the $y$-axis.}
\label{f:approxGP}
\end{figure}
Figure \ref{f:approxGP} summarizes the result of an experiment utilizing
up to 16 cores and 2 GPUs.  Problem sizes are varied along the columns of the
figure ($(n=50,N'=1000)$, $(n=50, N'=2000)$, $(n=128,N'=2000)$) and a number
of {\tt OpenMP} CPU threads and GPUs is varied in each plot. The top row in
the figure shows log timings, whereas the bottom row shows times relative to
the CPU-only calculation, via the same ratio used in Figure \ref{f:onecpu}.

First consider the results obtained for a single CPU thread.  The
figure shows that the ALC GPU computation yields between a 6x and 20x speedup,
with the better ratios obtained as the fidelity of the approximation
increases.  Note that the 1-GPU and 2-GPU results are the same in this case---a 
single CPU thread cannot make effective use of multiple GPUs.

When multiple {\tt OpenMP} CPU threads are in use the compute times steadily
decrease as more processors are added.  For example, considering the CPU-only
results, the ratio of the 1-CPU time to the 16-core CPU time is 14.5 for all
three problem sizes, which suggests very good efficiency (the best we could
hope for is 16x).  In the case of the first column, the 16 CPU thread solution
is faster than the 1-CPU-thread version interfacing with a GPU.  (In fact it
is better up to about 5 CPU threads for the GPU). Using 16 CPU threads
performing CPU-only calculations is almost as good as allowing those same 16
threads to queue jobs on a single GPU, which is creating a bottleneck.  The
results for two GPUs are much better, and we would expect the relative timings
to be even better with more GPUs.

We conclude that both SMP and GPU paradigms are helpful for calculating the
local GP approximation.   Their combined efforts, compared to using a single
CPU alone, represent a 33x, 50x, and 100x, improvement in wall-clock time on
the three problem sizes respectively.  If only one option were made available,
single CPU-only or single GPU+CPU, the latter is clearly preferable (giving
6-20x speedups).  However, adding on multiple CPU threads can lead to a 2.6x
speedup in the 1-GPU setting, and about 5x in the 2-GPU case, for all three
problems. Finally, augmenting a 16 CPU-only setup with 2 GPUs gives a 2.2x,
3.3x, 6.7x speedups respectively.  

We noticed in this latter case, with 16 CPUs and 2 GPUs, that only 5/16 of CPU
capability was being utilized with so much of the computation being off-loaded
to the GPUs.  We found that when this happens it is possible to get a further
$1/4$ reduction in wall-clock times by creating new CPU threads (10 or so) to
do CPU-only ALC calculations alongside the GPU ones.  Some pilot tuning is
needed to get the load balancing of CPU v.~GPU calculations right. Figure
\ref{f:approxGP} can serve as a guide, starting with an 80/20 GPU/CPU split
for the lowest fidelity case, increasing 85/15 and 90/10 as the fidelity, and
thus relative speedup obtained from the GPU, is increased.

\section{Big computer emulation}
\label{sec:big}

Here we demonstrate a three-level cascade of parallelism towards approximate
GP emulation on very big computer experiment data.  The first two levels are
{\tt OpenMP}-based and {\tt CUDA}-based, using CPUs and GPUs respectively, on
a single compute node. The third level is a cluster, allowing simultaneous use
of multiple nodes.  For this we use the simple network of workstations model
implemented in the {\tt snow} package
\citep{snow} for {\sf R}, which only requires a simple wrapper function
to break up the predictive locations $\mathcal{X}$ into chunks--- allowing
each to be processed on a separate node via {\tt clusterApply}---and then
to combine the outputs into a single object.\footnote{The built-in {\tt parallel}
package can also be used instead of {\tt snow}.  In our setup, cluster nodes
are allocated via {\tt SLURM}, an open-source Linux scheduler, and are
connected by an Infiniband FRD10 fabric.}

We remark that in the case of a single node with multiple cores, our use of
{\tt snow} accomplishes something very similar to an {\tt OpenMP} SMP
parallelization. However, given a choice the latter is faster than the former
since establishing a cluster requires starting multiple copies of {\sf R},
sending copies of the data to each, and then combining the results.  
Also, our
focus here is primarily on timing results, reminding readers that
fidelity/computational demands are tightly linked with emulation accuracy.  In
the case of our second example, \cite{gramacy:apley:2014} already illustrated
how a relatively thrifty approximation can provide more accurate
predictions compared to modern alternatives in a fraction of the time.

\subsection{Langley Glide-Back Booster}
\label{sec:lgbb}

Our first example is a real computer experiment for a re-usable rocket booster
called the Langley Glide-Back Booster.  The computer model, developed at NASA,
involves computational fluid dynamics (CFD) codes that simulate the
characteristics of the booster as it re-enters the atmosphere---modeling
outputs such as lift as a function of inputs such as speed, angle of attack,
and side-slip angle.  For more details of the experiment, including how the
emulator can benefit from a nonstationary/localized modeling capability due to
abrupt dynamical transitions at speeds near the sound barrier, see
\cite{gra:lee:2009}.    The design\footnote{The version of the data we consider 
here is actually the
output of Gramacy \& Lee's emulation of computer simulations adaptively
designed to concentrate more runs near speeds of Mach one, i.e., at the sound
barrier, using a partition-based non-stationary model.} has $N=37908$
3-dimensional input configurations, and six outputs but we only consider the
first one, lift, here.  The design is gridded to be dense in the first
input, speed, and coarse in the last, side slip angle.  We
consider using the local GP approximation method to interpolate the lift
response onto a regular grid that is two-times more dense in the first input,
and three times more dense in the second two.  That gives a predictive grid of
size $|\mathcal{X}| = 644436$. 

Our setup here mimics the apparatus described in Section
\ref{sec:fullapprox}, using 4 identical compute nodes, each
having 16 cores and 2 GPUs. We establish 16 {\tt OpenMP} CPU threads queuing
GPU ALC calculations on both GPUs, and 12 further {\tt OpenMP} CPU threads
performing CPU-only ALC calculations, initially allowing the GPUs to take 80\%
of the ALC work.  The {\tt snow} package distributes an equal workload to each
of the four nodes.  The ALC searches are over $N' = 1000$ NN candidate
locations, starting at $n_0=6$ and ending at $n=50$, i.e., following the
left-most panel in Figure \ref{f:approxGP}.  The wall-clock time for the full
emulation was 21 minutes. By way of comparison, a single CPU-only version (but
fully utilizing its 16 cores) takes 235 minutes (4 hours) and fully using all
CPU cores on all four nodes takes about 58 minutes (1 hour). Therefore the
GPUs yield about a 4x speedup, which is a little better than the 2x speedup
indicated in the bottom-left panel Figure
\ref{f:approxGP}.  

A higher fidelity search with $N' = 2000$, mimicking the middle panel of
Figure \ref{f:approxGP} except the CPU/GPU load was beneficially re-balanced
so that GPUs do 90\% of the work, took 33 minutes in the full (4x
2-GPU/16-CPU) setting.  A single CPU-only version (16 cores) takes 458 minutes
($\sim$8 hours) and using all four nodes takes 115 minutes ($\sim$2 hours). 
So the GPUs yield a 4x speedup, which is in line with the bottom-middle
panel of Figure \ref{f:approxGP}.
Increasing the fidelity again to $N' = 10$K, and re-balancing the load so that
GPUs do 95\% of the ALC work by allocating 12 extra CPU threads to do the
rest, takes 112 minutes on all 4x 2-GPU/16-CPU nodes, representing a more than
5x speed-up compared to the 4x 16-CPU (i.e., no GPU) version.  Alternatively,
keeping $N'=2000$ but increasing the local design size to $n=128$, mimicking
the right column of Figure \ref{f:approxGP} except with GPUs again doing 95\%
of the ALC work, takes 190 minutes, representing an almost 6x speedup.

\subsection{A one-hour supercomputing budget}
\label{sec:hour}

We wrap up with a search for the largest emulation possible on the resources
available to us. For data generation we chose the borehole function
\citep{worley:1987,morris:mitchell:ylvisaker:1993} which
%, although too oft used in the literature, makes for 
provides a familiar benchmark.  It has an 8-dimensional input space, and our
use of it here follows directly from \cite{gramacy:apley:2014} who copied the
setup of \cite{kaufman:etal:2012}; more details can be found therein.

Table \ref{t:borehole} summarizes the timings and accuracies of designs from
size $N=1000$ to just over $N=1$M, stepping by factors of two.  To keep things
simple, the predictive set size is taken to match the design size
($|\mathcal{X}| = N$), but note that they are different random (Latin
hypercube) samples. We allowed the fidelity of the approximation to increase
with $N$ along a schedule that closely matches settings that have worked on
similarly sized problems.  Specifically, we started with $(n=40, N'=1000)$ for
the smallest problem ($N=1000$), and each time $N$ doubled we increased $n$
additively by two and $N'$ multiplicatively by $1.5$.  The left panel of the
table shows results from a 96-node CPU cluster, where each node has 16 cores.
The middle panel shows results from a 5-node GPU/CPU cluster, where each node
has 2 GPUS and 16 cores.  Unfortunately, the infrastructure we had access to
did not allow CPU and GPU/CPU nodes to be mixed.  The final panel shows the
speedup-factor from the GPU nodes assuming we had 96 instead of five.

\begin{table}[ht!]
\centering
\begin{tabular}{rrr|rrr}
    &     &      &  \multicolumn{2}{c}{96x CPU}\\
\hline
$N$ & $n$ & $N'$ & seconds & mse \\ 
\hline
1000 & 40 & 100 & 0.48 & 4.88 \\ 
2000 & 42 & 150 & 0.66 & 3.67 \\ 
4000 & 44 & 225 & 0.87 & 2.35 \\ 
8000 & 46 & 338 & 1.82 & 1.73 \\ 
16000 & 48 & 507 & 4.01 & 1.25 \\ 
32000 & 50 & 760 & 10.02 & 1.01 \\ 
64000 & 52 & 1140 & 28.17 & 0.78 \\ 
128000 & 54 & 1710 & 84.00 & 0.60 \\ 
256000 & 56 & 2565 & 261.90 & 0.46 \\ 
512000 & 58 & 3848 & 836.00 & 0.35 \\ 
1024000 & 60 & 5772 & 2789.81 & 0.26 \\ 
\hline
\end{tabular}
\hspace{1cm}
\begin{tabular}{rr}
\multicolumn{2}{c}{5x 2 GPUs}\\
\hline
seconds & mse \\ 
  \hline
1.95 & 4.63 \\ 
2.96 & 3.93 \\ 
5.99 & 2.31 \\ 
13.09 & 1.74 \\ 
29.48 & 1.28 \\ 
67.08 & 1.00 \\ 
164.27 & 0.76 \\ 
443.70 & 0.60 \\ 
1254.63 & 0.46 \\ 
4015.12 & 0.36 \\ 
13694.48 & 0.27 \\ 
\hline
\end{tabular}
\hspace{1cm}
\begin{tabular}{r}
$\frac{\mathrm{CPU}}{5\cdot\mathrm{GPU}/96}$\\
  \hline
efficiency \\
  \hline
4.73 \\ 
4.26 \\ 
2.79 \\ 
2.66 \\ 
2.61 \\ 
2.87 \\ 
3.29 \\ 
3.63 \\ 
4.01 \\ 
4.00 \\ 
3.91 \\
   \hline
\end{tabular}
\caption{Timings and out-of-sample accuracy measures for increasing problem
sizes on the borehole data. The ``mse'' columns are mean-squared predictive
error to the true outputs on the $|\mathcal{X}| = N$ locations from separate runs 
(hence the small discrepancies between the two columns).  Both CPU and
GPU nodes have 16 CPU cores.  So the ``96x CPU'' shorthand in the table indicates 
1536 CPU cores.}
\label{t:borehole}
\end{table}

%We observe the following from the tables. 
On the CPU nodes, over a million
inputs and outputs can be processed in under an hour, $\sim100$K in about a
minute, and $\sim 10$K in about two seconds.  The GPU/CPU cluster has a little
less than half of the capacity, processing half-a-million points in just over
an hour.   Assuming we had more GPU/CPU nodes, the final column suggests that
the GPUs make the whole execution 2.5-4.5x faster on these problems. Notice
that these efficiencies decrease and then increase again as fidelity is
increased.  The initial high efficiencies are actually due to inefficiencies
in the {\tt snow} execution: 96 cores is overkill for problems sized in the
few thousands, requiring too much of a communication overhead between master
and slave nodes.  Whereas the latter high efficiencies are due to improvements
in GPU throughput for larger problems. Finally, notice that accuracy
(out-of-sample MSE) is steadily improving as fidelity increases. By way of
comparison,
\cite{gramacy:apley:2014} showed that with $N\in\{4000,8000\}$ the
approximations were at least as accurate as those in
\cite{kaufman:etal:2012} with less than 1\% of the computing effort.

However, comparisons based on accuracy in this context are at best
strained. In cases when each method can execute fast enough to perform a full
analysis (e.g., limiting to $n<10000$), we've observed (based on comparisons
like the ones above) accuracies that are strikingly similar
across a wide swath of comparators. We think it is reasonable to suggest that
would remain true for larger problem sizes, although this is nearly
impossible to verify. Often the largest runs reported by authors are on
proprietary data, and some involve proprietary library routines, which makes
reproducibility difficult.  For example, the largest problem entertained by
\cite{kaufman:etal:2012} was a cosmology example with $(N=20000,
|\mathcal{X}|=80000)$, but timing information was not provided and the
data is not publicly available to our knowledge.  
\cite{paciorek:etal:2013} entertained $(N=67275, |\mathcal{X}|=55379)$, but 
again without timing information or public data. Therefore, we conclude that
our method is at worst a worthy competitor relative to these alternatives, but
offering the potential for similar emulation quality on problems that are
several orders of magnitude larger.

\section{Discussion}
\label{sec:discuss}

The local GP approximation of \cite{gramacy:apley:2014} swaps a large problem
for  many small independent ones. We show in this paper how those many small
problems can be solved on a cascade of modern processing units.  We think this
is particularly timely research.  Many modern desktops have multiple cores and
(sometimes) multiple GPUs, and many modern ``supercomputers'' are not much
more than enormous clusters of high-end multi-core desktops and
GPUs.\footnote{Some really modern supercomuters are essentially clusters of
GPUs, with very little CPU computing capability, although we did not have
access to such a setup for the empirical work in this paper.}  Our primary
focus was on a GPU accelerated version of a key subroutine in the
approximation, allowing a faster execution at lower cost. Although results
have emerged casting doubt on some of the speed claims made in scientific
computing contexts for GPUs \citep[e.g.,][]{Lee:etal:2010}, it is still the
case that, penny-for-flop, GPUs are cheap.  Therefore, its proportion of
available flops will continue to grow relative to CPUs for some time to come.

We take a different tack to the use of GPUs for GP computer emulation compared
to other recent works, which primary offload large matrix
calculation to GPUs.  As we show, the combined effects of approximation and
massive parallelization can extend GP emulation to problems at least
an order of magnitude larger than what is currently possible.  We note that
others have had similar success parallelizing non-GP models for
computer emulation. For example, \cite{pratola:etal:2013}  parallelized
the Bayesian additive regression trees (BART) method using the message passing
interface (MPI) and report handling designs as large as $N=7$M using hundreds
of computing cores. Such efforts will likely remain in vogue so
long as computing resources continue to grow ``out'' (with more nodes/cores, etc.)
faster than they grow ``up'', which will be for quite some time to come.

\subsection*{Acknowledgments}

This work was completed in part with resources provided by the University of
Chicago Research Computing Center.  Many thanks to Matt Pratola for comments
on an early version.  We are grateful for valuable comments from two referees and
an associate editor during the formal review process.

\appendix

\section{Double reduction {\tt CUDA} code}
\label{sec:sum2}

The {\tt CUDA} GPU kernel {\tt sumBoth} assumes that both inputs, \verb!d_data1! and
\verb!d_data2! have length \verb!n!, and the call is
\verb!sumBoth<<1,n>>(d_data1, d_data2)! so that \verb!blockDim.x = n!.

{\singlespacing
\begin{verbatim}
__global__ void sumBoth(double *d_data1, double *d_data2)
{
  int tid = threadIdx.x;
  int nelem = blockDim.x;
  int nTotalThreads = NearestPowerOf2(nelem);
  int halfPoint = (nTotalThreads >> 1);

  if (tid < halfPoint) {
    int thread2 = tid + halfPoint;
    if (thread2 < nelem) {
      d_data1[tid] += d_data1[thread2];
      d_data2[tid] += d_data2[thread2];
    }
  }
  __syncthreads();

  // now its a regular power of 2 reduction on data of size halfPoint
  if (tid < halfPoint){
    if(tid < halfPoint/2) { // First 1.2 of the threads work on d_data1
      for(unsigned int s=halfPoint/2; s>0; s>>=1) {
        if (tid < s)  d_data1[tid] += d_data1[tid + s];
         __syncthreads();
      }
    } else { // Second 1/2 of the threads works on d_data2
      tid = tid - (halfPoint/2);
      for(unsigned int s=halfPoint/2; s>0; s>>=1) {
      if (tid < s) d_data2[tid] += d_data2[tid + s];
      __syncthreads();
    }
  }
}
\end{verbatim}}

\bibliography{../approx_gp,gpu}
\bibliographystyle{jasa}

\end{document}